\documentclass[preprint,prd,tightenlines,nofootinbib,superscriptaddress]{revtex4-1}
\usepackage{color}
\usepackage{graphicx}
\definecolor{red}{rgb}{1,0,0}
\def\lesssim{\ \hbox{\raise 2pt \hbox{$<$} \kern -13pt
                     \lower 3pt \hbox{$\sim$}}\ }
\def\greatersim{\ \hbox{\raise 2pt \hbox{$>$} \kern -13pt
                     \lower 3pt \hbox{$\sim$}}\ }

\def\pythia{{\sc Pythia}}

\def\powheg{{\sc Powheg}}

\input epsf.tex
\def\desepsf(#1 width #2){\epsfxsize=#2 \epsfbox{#1}}
\def\kt{\ensuremath{k_t}}

\newcommand{\Pmax}{p}
\newcommand{\cA}{{\cal A}}

\usepackage{amsmath,bm}

\begin{document}
\hspace*{12.9 cm} {\small DESY 14-085} 
\vspace*{1.4 cm} 
\title{Hadroproduction of 
electroweak gauge boson plus jets 
 and TMD parton density functions}
\author{S.\ Dooling} 
\affiliation{Deutsches Elektronen Synchrotron, D-22603 Hamburg}
\author{F.\ Hautmann} 
\affiliation{Dept.\  of Physics and Astronomy, 
University of   Sussex,   Brighton   BN1  9QH}
\affiliation{Rutherford Appleton Laboratory,  Chilton  OX11  0QX}
\affiliation{Dept.\  of  Theoretical Physics, 
University of Oxford,    Oxford OX1 3NP}  
\author{H.\ Jung}
\affiliation{Deutsches Elektronen Synchrotron, D-22603 Hamburg}
\affiliation{Elementaire Deeltjes Fysica, Universiteit Antwerpen, B 2020 Antwerpen}
\begin{abstract}
If studies of  electroweak gauge boson final states 
at the Large Hadron Collider,  
for  Standard Model physics and beyond,  
are sensitive to effects of  the initial state's 
 transverse momentum distribution,   appropriate 
generalizations of  QCD shower evolution 
 are required.  We  propose a  method to do this  based on 
QCD  transverse momentum dependent (TMD) 
  factorization  at high energy. The method  incorporates  
  experimental information from the   
high-precision deep inelastic scattering (DIS) measurements,  
and includes experimental and theoretical uncertainties on 
TMD parton density functions.  We illustrate the approach  
 presenting   results for  production of 
$W$-boson + $n$ jets at the LHC, including 
azimuthal correlations  and 
subleading jet distributions. 
\end{abstract} 


\maketitle

The associated production of an electroweak gauge boson and 
 hadronic  jets is central to many aspects of the Large Hadron 
Collider (LHC) physics program. It is an important background to 
Higgs boson and top quark studies, and to supersymmetry and  
dark matter  searches~\cite{cigdem}.  It provides 
benchmark observables 
  for studies of  QCD, 
Monte Carlo event generators and  parton density 
functions~\cite{pdf4lhc-alekhin}. 
In the  upcoming high-luminosity runs,  
it can be used in combination  with Higgs boson 
production~\cite{cipriano,dendunnen} for  precision 
 studies of QCD initial-state effects beyond fixed-order 
perturbation theory. 

Baseline predictions   
are  obtained   from  
next-to-leading-order (NLO) 
perturbative matrix elements for the hard, high-$p_\perp$ 
process, 
 matched with parton 
 showers describing the collinear evolution of the jets 
developing from the hard event~\cite{hoeche13}. 
When    this  perturbative QCD 
 picture  is pushed  to higher and higher
 energies $\sqrt{s}$, however,    new effects 
 arise in  the 
jet multiplicity distributions  and 
   the structure of  angular correlations,  
  due to soft  but       finite-angle  
   multi-gluon 
emission. As  was noted already long ago~\cite{mw92},  
 these  high-energy effects   can be taken into account   by  
treating the QCD evolution of the  initial-state parton 
distributions  via  
     transverse-momentum 
dependent  branching algorithms  
  coupled~\cite{hef}  to hard matrix 
elements at  fixed transverse momentum.  
This allows one to include   soft gluon 
coherence~\cite{Catani:1989sg}     
not only for 
collinear-ordered emissions but also 
 in the non-ordered region that 
opens up at high $\sqrt{s} / p_\perp$ and large $p_\perp$. 
(Examples of  angular  correlations  in multi-jet  deep inelastic 
scattering (DIS)  final states 
are studied in~\cite{hj-ang}. 
 See e.g.~\cite{fh09-highmul}  and references therein.) 

Besides these  dynamical effects,      
the role of 
including 
 the correct  transverse-momentum kinematics  in 
 branching algorithms describing  
QCD evolution in  Monte  Carlo 
event generators   
has recently been  
emphasized in~\cite{sama1,sama2},  and connected 
with  experimental observations of  $p_\perp$ spectra  
 at the LHC~\cite{jetlhc-refs}     
in the case of        
jets produced  at moderately  
 non-central  rapidities.   
It has  been pointed out~\cite{sama1,sama2} 
 that collinear approximations, combined with  
energy-momentum conservation constraints,  
give rise to non-negligible kinematic 
 shifts   in longitudinal 
momentum distributions, and   
are responsible for   a large fraction  of  
parton showering 
 corrections to LHC jet final states~\cite{jetlhc-refs}. 

  In this paper we propose an approach  
to  electroweak boson plus jets 
production which  addresses both the   
 dynamical and kinematical issues   mentioned above  
 via transverse-momentum dependent (TMD) 
QCD evolution equations, and corresponding parton density functions 
and perturbative matrix elements.  
Traditional approaches to 
electroweak  boson production 
taking into account the initial state's 
transverse momentum distribution have focused on 
the  boson  spectrum in the low-$p_\perp$ Sudakov 
  region,  
  and on  the   treatment  of    large 
 logarithms for transverse 
momenta small compared to the boson invariant mass. 
Our work  
 treats   physical effects which   persist 
at high $p_\perp$    and can affect 
 final states with high jet multiplicities.  
To this end    we  use  the transverse-momentum dependent 
QCD factorization~\cite{hef},  which is  valid  
   up to arbitrarily large  
$p_\perp$. We couple this with CCFM~\cite{Catani:1989sg}  
evolution equations for TMD 
gluon and valence quark densities using  the   
 results 
recently obtained  in~\cite{hj-updfs-1312}. 

This theoretical framework, 
although not limited in  $p_\perp$,    
 is  based on the high-energy 
expansion  $\sqrt{s}  \to \infty $.  
Non-asymptotic contributions are included through 
CCFM matching with soft-gluon   terms  in the  
evolution kernels  and  through  subleading effects  
in the flavor non-singlet sector    
 according to the method 
of~\cite{hj-updfs-1312}.  In~\cite{hj-updfs-1312}  
this approach is  applied to  deep inelastic 
scattering  (DIS) and 
charm quark production and confronted 
with high-precision 
combined HERA data~\cite{comb-charm,Aaron:2009aa}, which imply 
small longitudinal momentum fractions $x$.  
In contrast, 
the subject of this paper explores 
 processes which mostly  occur when  
the values of $x$ are not very small.  
 It   tests the  
matching procedure  and the 
 non-asymptotic contributions. By this calculation, we  
 push  the limits  of the high-energy expansion 
 beyond the small-$x$ region,  in a manner which can be 
controlled using the estimation of  
theoretical and experimental  uncertainties on 
  TMD  distributions  proposed   in~\cite{hj-updfs-1312}  within 
 the 
\verb+herafitter+   framework~\cite{Aaron:2009aa,herafitter}.  
  Given  the complexity of the final states considered, this is 
a challenging problem. 
The results  are  however encouraging.  Moreover,  they 
 are   sufficiently general  to be of interest to  
any  approach  that  employs   TMD formalisms in QCD  
to go beyond fixed-order perturbation theory and 
appropriately    take  account  of   nonperturbative effects. 
This  will be       relevant  both to  
precision  studies of Standard 
Model physics and to new physics searches 
for which gauge boson plus jets production is an important background.

Using the parton branching  Monte Carlo 
implementation   
of  TMD  evolution   developed in~\cite{hj-updfs-1312} 
we  make  predictions, including uncertainties, 
 for final-state observables associated with $W$-boson production.  
We study  jet  transverse momentum spectra and azimuthal correlations. 
 In particular,  we examine  subleading  jet distributions, measuring the 
transverse momentum imbalance between the vector boson and the 
leading jet. 

  The starting point of  our   approach  is to apply 
QCD high-energy factorization~\cite{hef}  at fixed transverse 
momentum to 
electroweak 
gauge boson + jet production, $ q + g^* \to V + q$, 
where $V$ denotes  a gauge boson and 
$g^*$ an  off-shell gluon.  The basic  observation is that this 
factorization allows one to sum 
 high-energy logarithmic corrections for 
$\sqrt{s} \to \infty$ to all orders in the QCD coupling provided 
the spacelike evolution of the off-shell gluon includes  
 the full BFKL anomalous dimension for 
longitudinal momentum fraction $x \to 0$~\cite{lipatov}. 
The   
CCFM evolution 
equation~\cite{Catani:1989sg}  is an exclusive branching 
 equation 
which satisfies  
this property.  
In addition,   it  includes  finite-$x $ 
contributions to parton splitting, incorporating  
soft-gluon coherence for any value of $x$. 
The evolution equation 
reads~\cite{Catani:1989sg,hj-ang}     
\begin{eqnarray}
\label{uglurepr1}
  {\cal A} ( x , \kt , p  ) & = & 
  {\cal A}_0 ( x , \kt , p  ) + 
\int { {dz} \over z} \int { { d q^2} \over q^2} \ 
\Theta   (p - z  q)  
\nonumber\\
& \times & 
 \Delta    (p , z  q) 
\ {\cal P} ( z, q, \kt)   
\   {\cal A} 
 ( { x \over z} , \kt  + (1-z) q, q )
 \hspace*{0.3 cm} ,        
\end{eqnarray} 
where $\cA(x,\kt,\Pmax)$ is the 
TMD gluon density function, 
depending on 
longitudinal momentum 
fraction $x$,  
transverse momentum 
$\kt$ and  evolution variable $\Pmax$. 
The first term in the right hand side of Eq.~(\ref{uglurepr1}) 
is the contribution of the 
non-resolvable branchings between  starting scale 
$q_0$ and   evolution scale $p$, while 
the integral term in the right hand side of Eq.~(\ref{uglurepr1}) 
gives the \kt-dependent branchings in terms of the 
 Sudakov form factor $\Delta$ and unintegrated 
  splitting function ${\cal P}$.  Unlike ordinary, integrated 
splitting functions, the latter encodes 
soft-virtual contributions into  the   
non-Sudakov  form 
factor~\cite{Catani:1989sg,hj-ang}.

In this framework the vector boson production cross section has the form 
\begin{equation}
\label{vector-fac} 
\sigma^{(V)} =  \int  {\cal A}  \otimes H_{qg} \otimes {\cal B}    
  \hspace*{0.3 cm} ,    
\end{equation}  
where the symbol $\otimes$ denotes convolution in both longitudinal and transverse 
momenta, 
${\cal A}$ is the gluon density  function obeying Eq.~(\ref{uglurepr1}), 
$H$ is the off-shell  (but gauge-invariant) 
 continuation of the $ q g$ hard-scattering function  
specified by the high-energy  factorization~\cite{hef}, 
and ${\cal B}$  is 
the valence quark density function 
introduced at unintegrated level 
according to the 
method~\cite{Deak:2010gk}, such that it obeys a  
modified CCFM branching equation.   
Explicit calculations for $H$ are carried out 
 in~\cite{marball,hent12,vanham,baranov} with off-shell 
partons~\cite{fad06,lip01}.\footnote{Ref.~\cite{hej} 
provides an approach to vector boson plus jets  also inspired 
by   QCD high-energy factorization~\cite{hef}. 
 This approach differs 
from that of the present paper as it is based on matching 
tree-level $n$-parton amplitudes with  BFKL amplitudes in the 
 multi-Regge kinematics,   treating initial-state partons  as collinear. 
 TMD parton density functions  and 
 \kt-dependent branching evolution do not enter in the approach~\cite{hej}.}    

The  $ {\cal A}_0 $ 
term  in the right hand side of Eq.~(\ref{uglurepr1}), and  the 
analogous term  in the  modified CCFM branching  equation
for the  quark  distribution ${\cal B}$~\cite{Deak:2010gk}, 
depend on  nonperturbative parton  distributions  
at scale $q_0$, which are 
to be determined from fits to experimental data.   
We here use the determination~\cite{hj-updfs-1312}  
 from  the  precision 
measurements of the
$F_2$  structure function~\cite{Aaron:2009aa}   
in the range  $x<0.005$, $Q^2>5$~GeV$^2$,  and 
the precision  measurements of  the charm  structure function 
$F_2^{({\rm{charm}})} $~\cite{comb-charm} in the range 
$Q^2 > 2.5$~GeV$^2$.   
Good fits to $F_2$ and $F_2^{({\rm{charm}})} $ are obtained    
(with the best fit to 
$F_2^{({\rm{charm}})} $ 
giving  
$\chi^2$ per degree of freedom 
$\chi^2/ndf \simeq 0.63$, and the best fit to 
$F_2$  giving  $\chi^2/ndf \simeq 1.18$~\cite{hj-updfs-1312}). 
Despite the   limited kinematic range,  
 the great  precision of  the combined 
data~\cite{Aaron:2009aa,comb-charm}    provides 
    a  compelling test of the  approach  at small  $x$.   
The   production of   final states with 
$W$ boson and multiple  jets  at the LHC  receives 
contributions  from a non-negligible fraction of events with 
 large  separations in rapidity between final-state 
particles~\cite{wjet-dis13}, calling  for  parton 
branching  methods beyond the collinear 
approximation~\cite{mw92}.   On the other hand, the 
average values of longitudinal momentum fractions  $x$ at which  the gluon 
density is sampled in the 
 $ W  $-boson + jets cross sections at the LHC are not 
 very   small. Moreover, quark's average momentum fractions are moderate, 
and    quark  density   contributions 
  matter~\cite{hent12} at TMD level. 
For these reasons,   $ W  $ + jets   pushes   the limits of  the 
 approach   
probing it  in a region where  its  theoretical 
 uncertainties increase~\cite{hj-updfs-proc}, and where 
the DIS experimental 
data~\cite{Aaron:2009aa,comb-charm} do not  
constrain well the TMD gluon distribution.  

\begin{figure}[htb]
\includegraphics[scale=0.67]{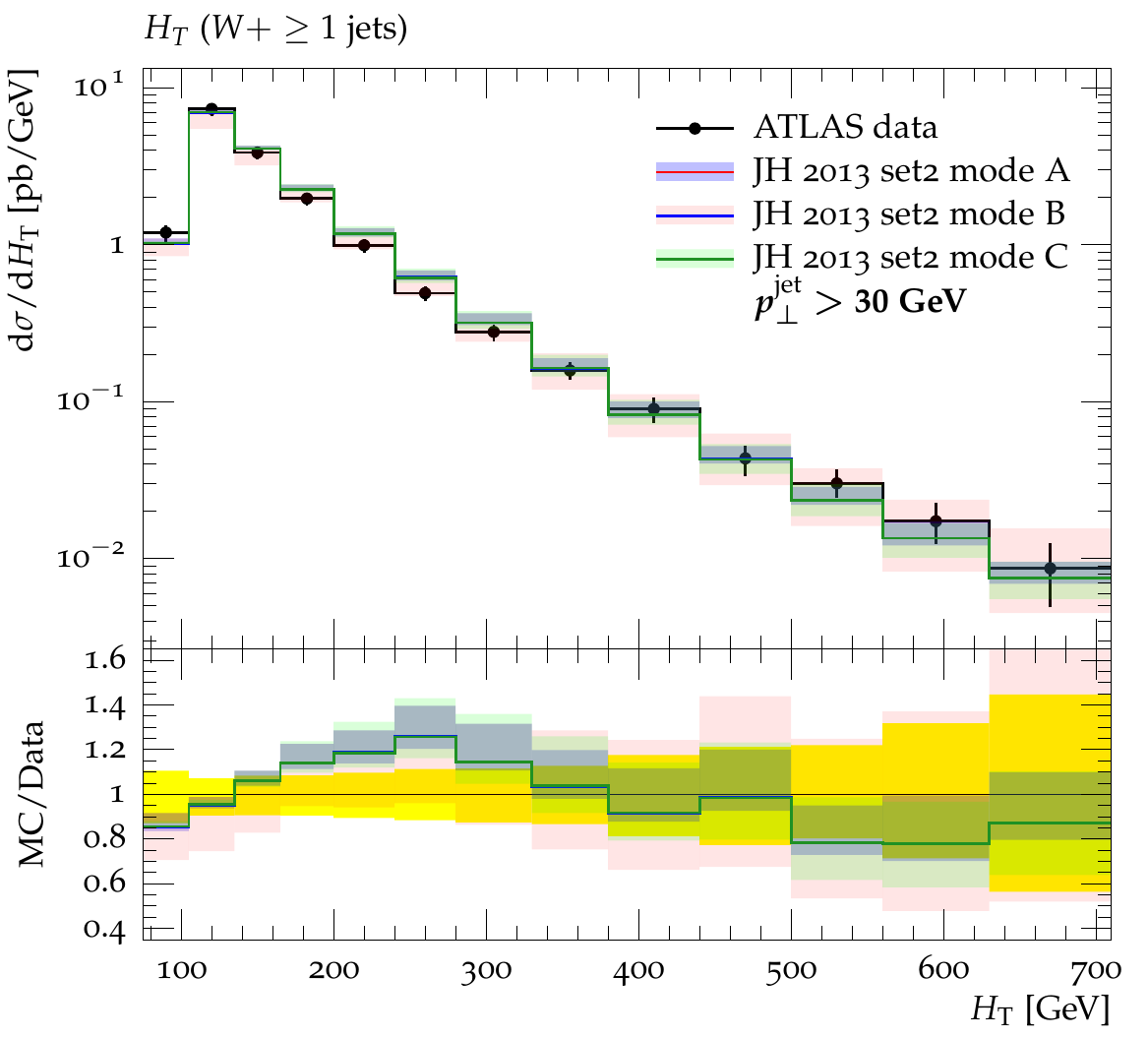}
\includegraphics[scale=0.67]{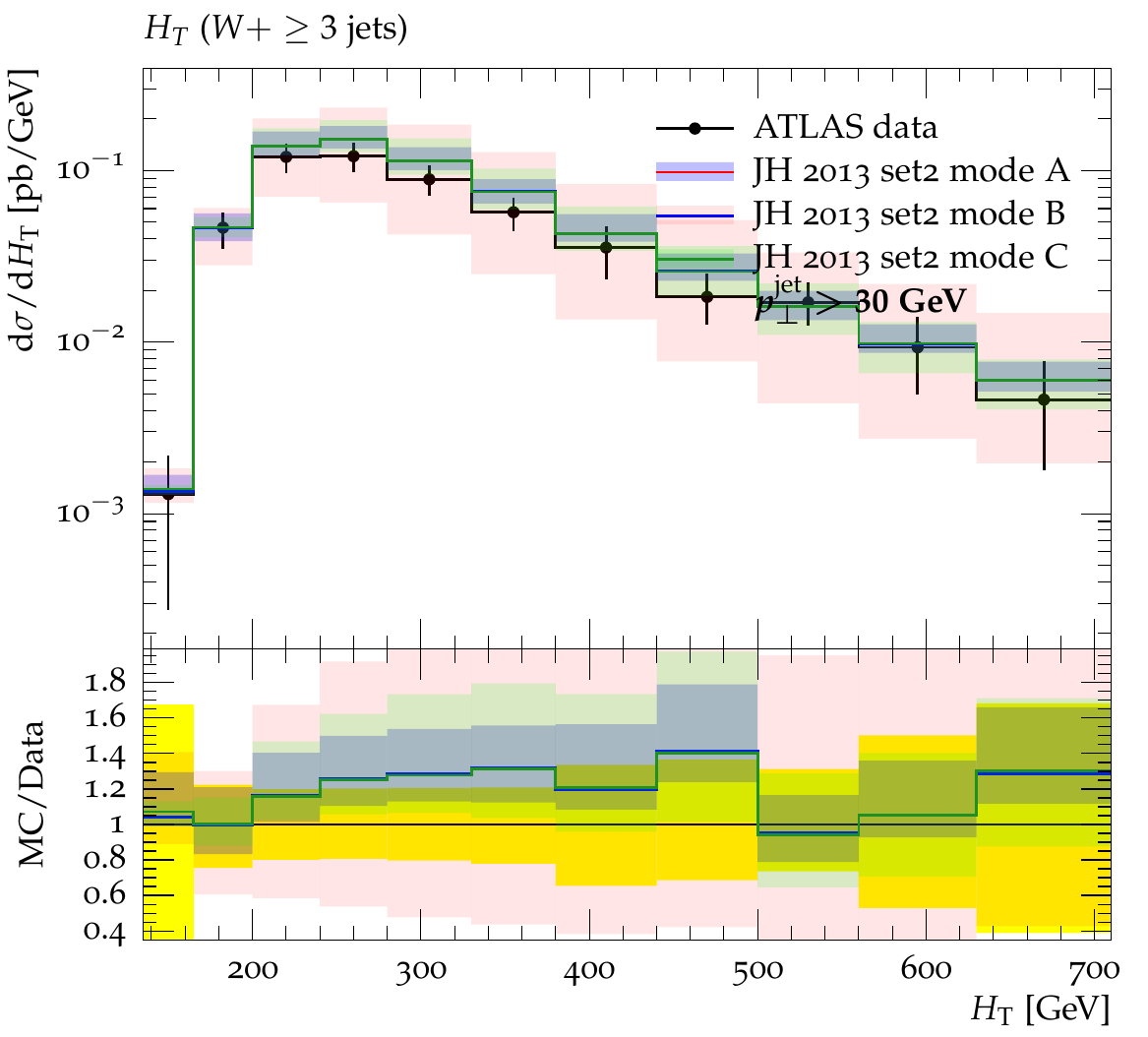}
  \caption{\it Total transverse energy $H_T$ distribution 
  in  final states   
with 
$W $-boson + $n$  jets  at the LHC, for 
(left)  $ n \geq 1 $,  
(right)  $n \geq 3$. 
The purple, pink and green bands 
correspond to mode A, mode B  and mode C 
as described in the text. 
 The experimental data are 
 from~\protect\cite{atlas-w-jets}, with the  
experimental uncertainty represented by  the yellow band.}
\label{fig:Ht}
\end{figure} 

 The numerical 
results  that follow are 
obtained using  the {\sc Rivet} - package~\cite{Buckley:2010ar}.  
We use the TMD distribution set JH-2013-set2~\cite{hj-updfs-1312}.   
We  compare  the results  
  with the ATLAS measurements~\cite{atlas-w-jets} 
(jet rapidity $|\eta |  < 4.4$)    and CMS 
measurements~\cite{cms-w-jets} (jet rapidity  $|\eta |  <  2.4 $).  
The uncertainties on the  predictions  are determined  according  
to the method~\cite{hj-updfs-1312}. This treats experimental and 
theoretical  uncertainties.  
Experimental pdf uncertainties are obtained 
  within the  \verb+herafitter+ package  following the 
procedure of~\cite{Pumplin:2001ct}. Theoretical uncertainties are 
considered separately 
due to the variation of the starting scale $q_0$ for 
evolution, the renormalization scale $\mu_r$ for 
the strong coupling, the factorization  scale $\mu_f$. 
We apply this method 
   in different   
modes: mode A (purple band in the plots) 
includes  uncertainties  due to the  
  renormalization scale, starting evolution scale,  
  and experimental errors;  mode B 
 (pink band in the plots) and mode C (green band in the plots) 
also  include  
 factorization scale uncertainties. These are estimated as follows.

\begin{figure}[htb]
\includegraphics[scale=0.67]{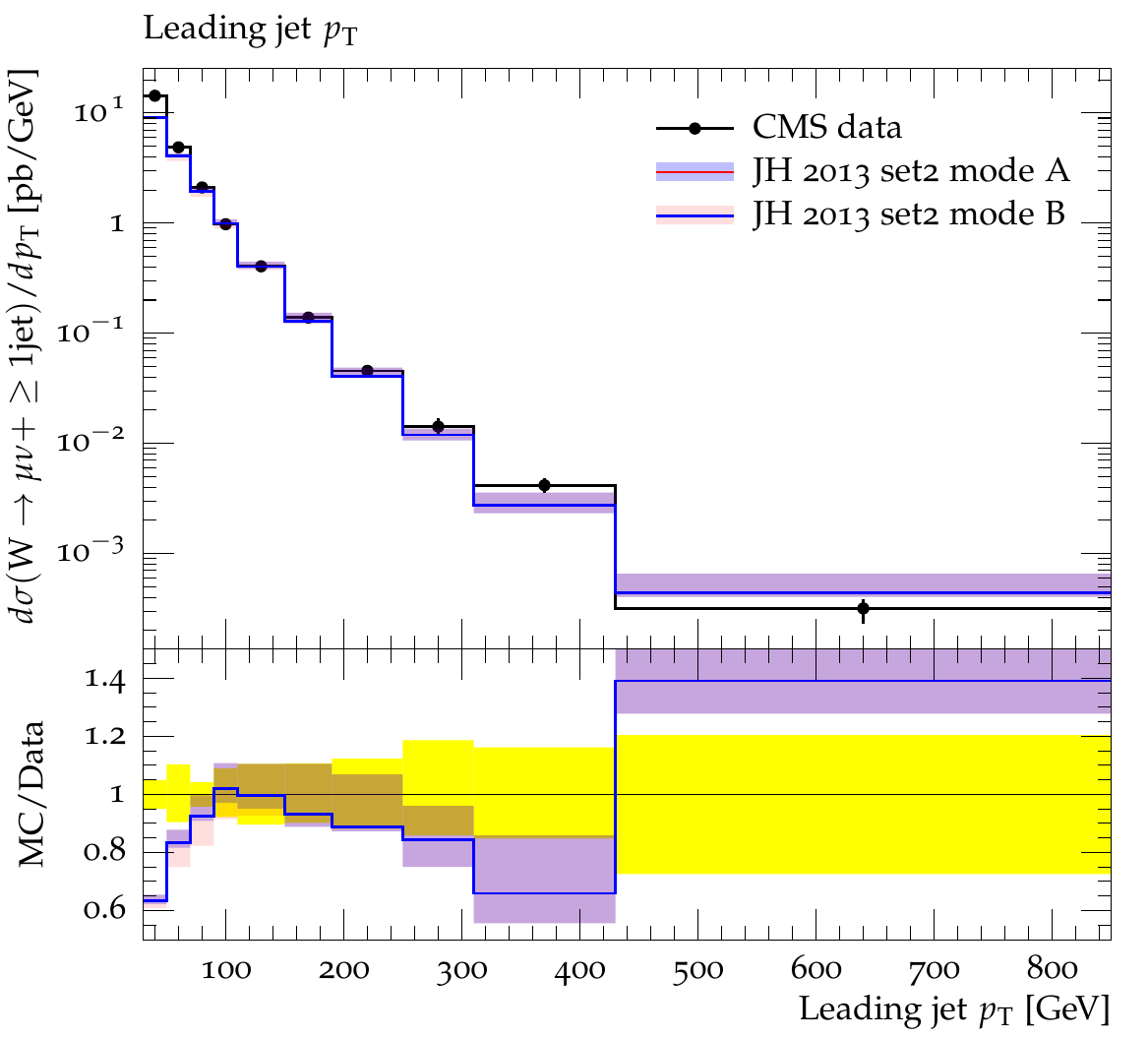}
\includegraphics[scale=0.67]{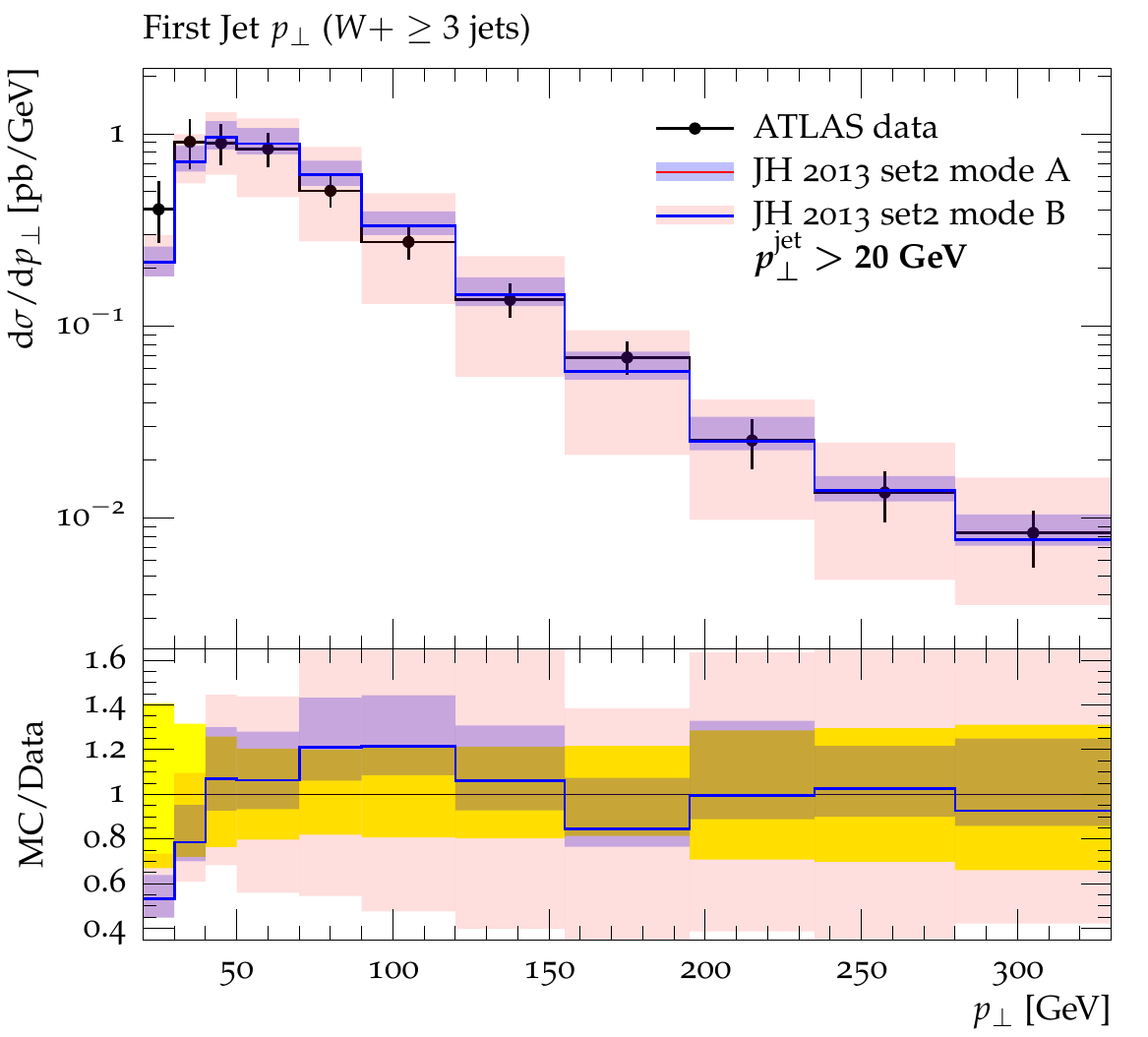}
  \caption{\it 
Leading jet $p_T$ spectra in   
$W $-boson + $n$  jets: (left) inclusive; (right)  $n \geq 3$. 
 The purple and pink bands 
correspond to mode A and mode B  
as described in the text. 
 The experimental data are  from~\protect\cite{cms-w-jets} (left) 
and~\protect\cite{atlas-w-jets} (right),  with the  
experimental uncertainty represented by  the yellow band.}
\label{fig:1stjet-pt}
\end{figure}

 We take the 
central value for the factorization scale to be 
 $\mu_f^2 = m^2 + q_\perp^2$, where $m$  and $q_\perp$ 
are the    invariant mass and transverse momentum of the boson + jet system. 
The choice of this scale is suggested by  the  
CCFM  angular ordering~\cite{mw92,Catani:1989sg,hj-ang}
and the maximum angle available to the branching. 
We then consider two different types of variation of $\mu_f$. 
In mode C, we vary the transverse  
part  of $\mu_f^2$ 
 by a  factor of 2 above and below the central value. In mode B, 
we decompose $\mu_f$ as  
 $ \mu_f^2 = m_V^2 + \nu^2$, where $m_V$ is the vector boson mass, and 
vary the dynamical   part $\nu^2$ of $\mu_f^2$, again  
 by a  factor of 2 above and below the central value.  
We note that 
the above variation affects the kinematics of the hard scatter, and 
 the amount of energy available for the shower.  
While the mode C variation is more closely related to the 
estimation of unknown higher-order corrections in standard   calculations performed 
under  collinear-ordering approximations, the mode B variation is a 
(conservative) way to estimate uncertainties from possibly enhanced 
higher orders due to  longitudinal momentum kinematics 
(not considered under standard approximations). 
For this reason we  expect  large mode-B uncertainties especially 
in the case of high  multiplicity. 
One of the limitations of  the current treatment is that 
this variation is applied to the shower but not to the hard matrix 
element. In a more complete calculation,   subject for future 
investigations, the scale dependence is taken into account  in the hard factor, and  
the pdf fitted to data is also changed~\cite{hj-updfs-1312}, unlike the ordinary case 
of collinear calculations. 
The net result of these two effects is expected to 
 reduce the uncertainty band. 
The present treatment, on the other hand, 
combined  with the sensitivity of the process to  the medium to large 
$x$ region,    leads to significant 
  theoretical uncertainties,    in particular  larger 
than the experimental  uncertainties. Thus, we  regard  
 the    mode B bands  presented in the following  
  as  the  most conservative estimate of the 
uncertainties. We expect mode C bands to be 
 smaller,  and  intermediate between mode A and mode B. 
 We note that  the factorization scale  variation 
 plays  a  different role here  than  in ordinary collinear calculations.

Fig.~\ref{fig:Ht} shows  the  total 
 transverse energy distribution $H_T$ 
for  production of  $W$-boson $ + n $ jets, 
 for different values of the number  of jets $n$. 
We take the minimum jet transverse momentum to be 30 GeV. 
 The  main  features of the final states are   described by the 
predictions  including the case of  higher   jet  multiplicities. 
The theoretical uncertainties are larger for larger $H_T$, 
 corresponding to increasing $x$. At fixed $H_T$, they are larger for higher 
jet multiplicities, corresponding to higher probability for jets to be formed from the 
partonic showers.  The comparison of the bands for the three modes 
described above 
 illustrates that mode C is intermediate between mode A and mode B.

 We next  consider    the   spectra of the individual jets. 
Fig.~\ref{fig:1stjet-pt} shows the  spectrum   of the leading 
jet associated 
with the $W$-boson, 
inclusively (left) and for $n \geq 3$ jets (right).  
For the sake of simplicity we only show uncertainty bands 
corresponding to the  two extreme cases, A and B (mode C is intermediate 
between these, similarly to the case of Fig.~\ref{fig:Ht}). 
The CMS~\cite{cms-w-jets} (left)  and 
ATLAS~\cite{atlas-w-jets}  (right) 
measurements  
 cover different ranges in jet rapidity, respectively  
$|\eta |  <  2.4 $~\cite{cms-w-jets}  and 
$|\eta |  <  4.4 $~\cite{atlas-w-jets}. 
The plot on the left  includes higher values of  $p_\perp$. 
Given the computational  limitations at finite $x$  outlined above, 
the theory comparison with the 
measurements in Fig.~\ref{fig:1stjet-pt}    
  is satisfactory 
over a  broad  $p_\perp$ range. 
\begin{figure}[htb]
\includegraphics[scale=0.67]{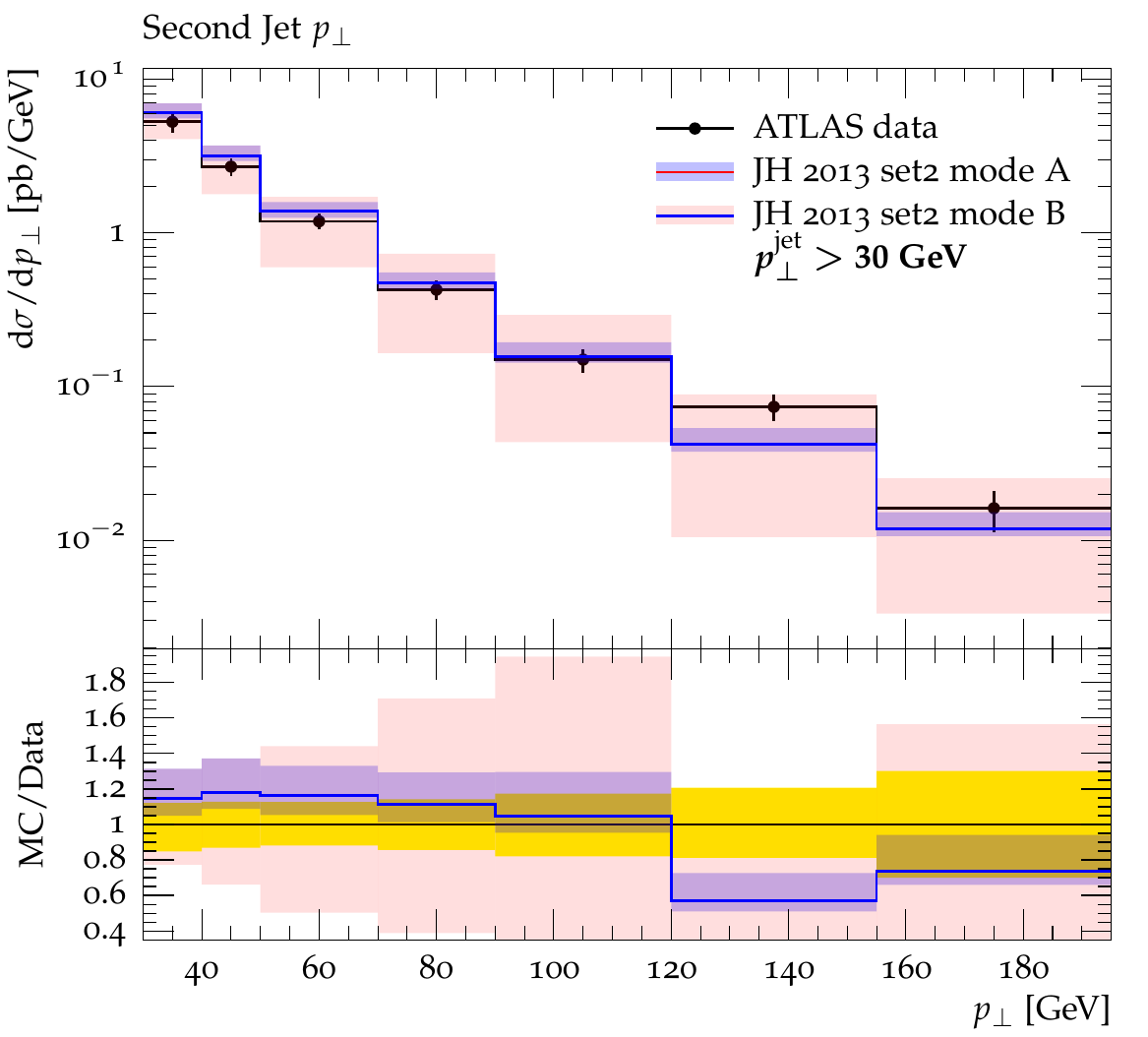}
\includegraphics[scale=0.67]{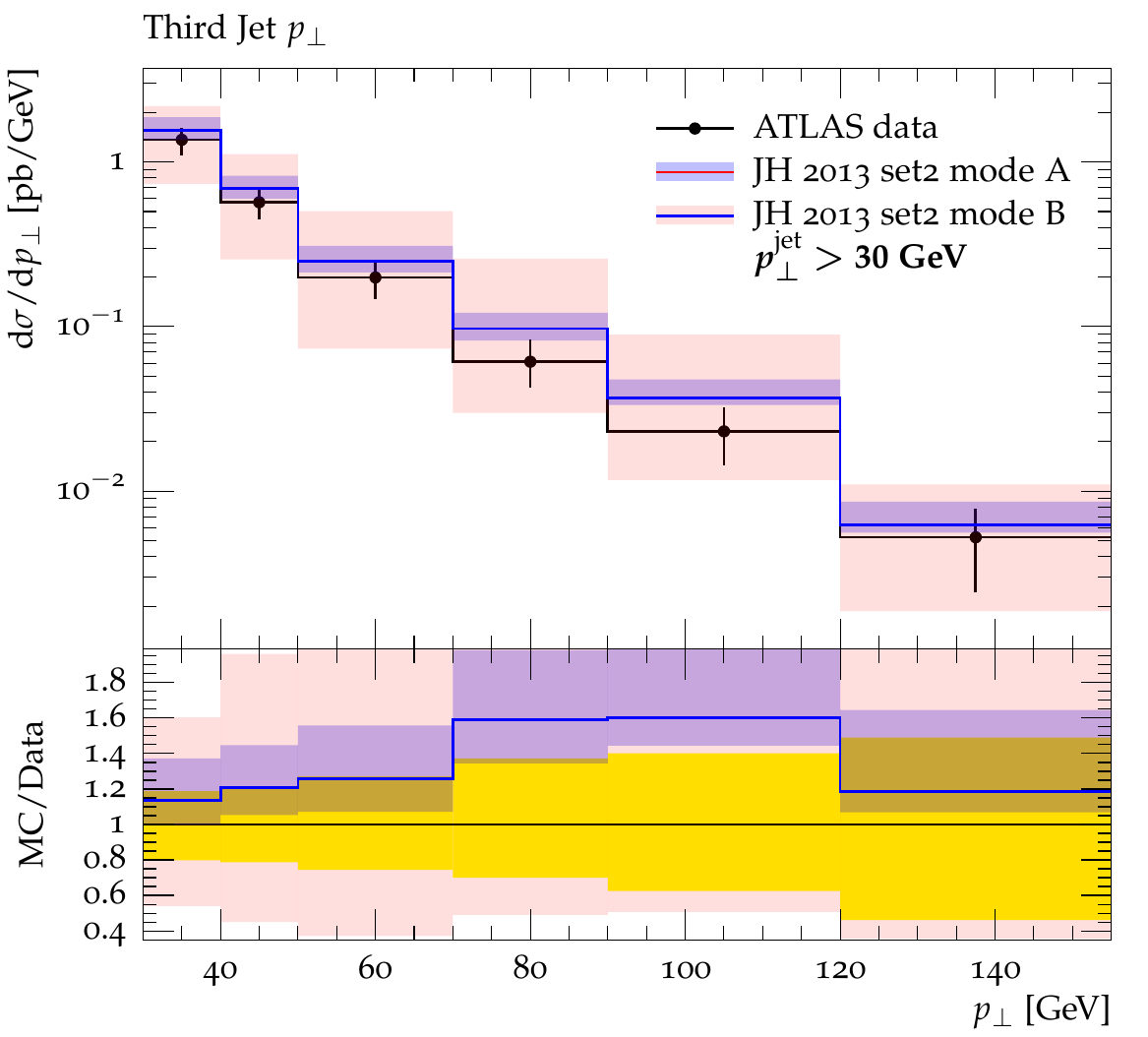}
  \caption{\it Second jet  (left)  
and third jet (right) distributions associated with   
$W $-bosons.    The purple and pink bands 
correspond to mode A and mode B  
as described in the text. 
 The experimental data are 
 from~\protect\cite{atlas-w-jets},  
with the  
experimental uncertainty represented by  the yellow band.}
\label{fig:3rdjet-pt}
\end{figure} 
 It is noted in~\cite{wjet-dis13}  
that, in contrast,    the 
leading-order 
\pythia~\cite{pythia8}  
result  strongly  deviates from these  measurements   
in the high-multiplicity and the 
  high-p$_\perp$ regions.   
In such a 
  framework  the description of the   
high-p$_\perp$ region  is to  be improved by 
supplementing the  parton shower 
with   next-to-leading-order corrections to the 
matrix element,  e.g.~via matched  
  NLO-shower  calculations~\cite{ma}  such as \powheg. 
The  TMD  formulation with exclusive evolution equations, 
  on the other hand, 
incorporating at the outset  large-angle, finite-k$_\perp$  
emissions~\cite{hj-ang,unint09}, 
  can   describe  the  shape of the spectra also at large 
multiplicity and large 
transverse momentum.  We note in particular that the different ranges in 
rapidity quoted above for the  samples~\cite{cms-w-jets,atlas-w-jets}    play 
a non-negligible role, 
given that our  exclusive formalism  is designed to 
treat gluon radiation over large rapidity intervals.

In 
Fig.~\ref{fig:3rdjet-pt}  we    look   into    the 
multi-jet  final states  
in closer  detail by  examining   the 
$p_\perp$ spectra of the    second jet  and  the third jet 
associated with $W$ production.  We  see 
  that not only the leading jet and  global 
distributions of  Figs.~\ref{fig:1stjet-pt}  
and~\ref{fig:Ht}   but also the detailed shapes 
of the subleading jets in 
Fig.~\ref{fig:3rdjet-pt}  
can be obtained from  the  TMD  formalism. 
The uncertainty bands,  on 
the other hand, increase as we go to higher jet 
multiplicity. The effect is moderate 
for mode A, but   pronounced  for the conservative 
 mode B.

\begin{figure}[htb]
\includegraphics[scale=0.67]{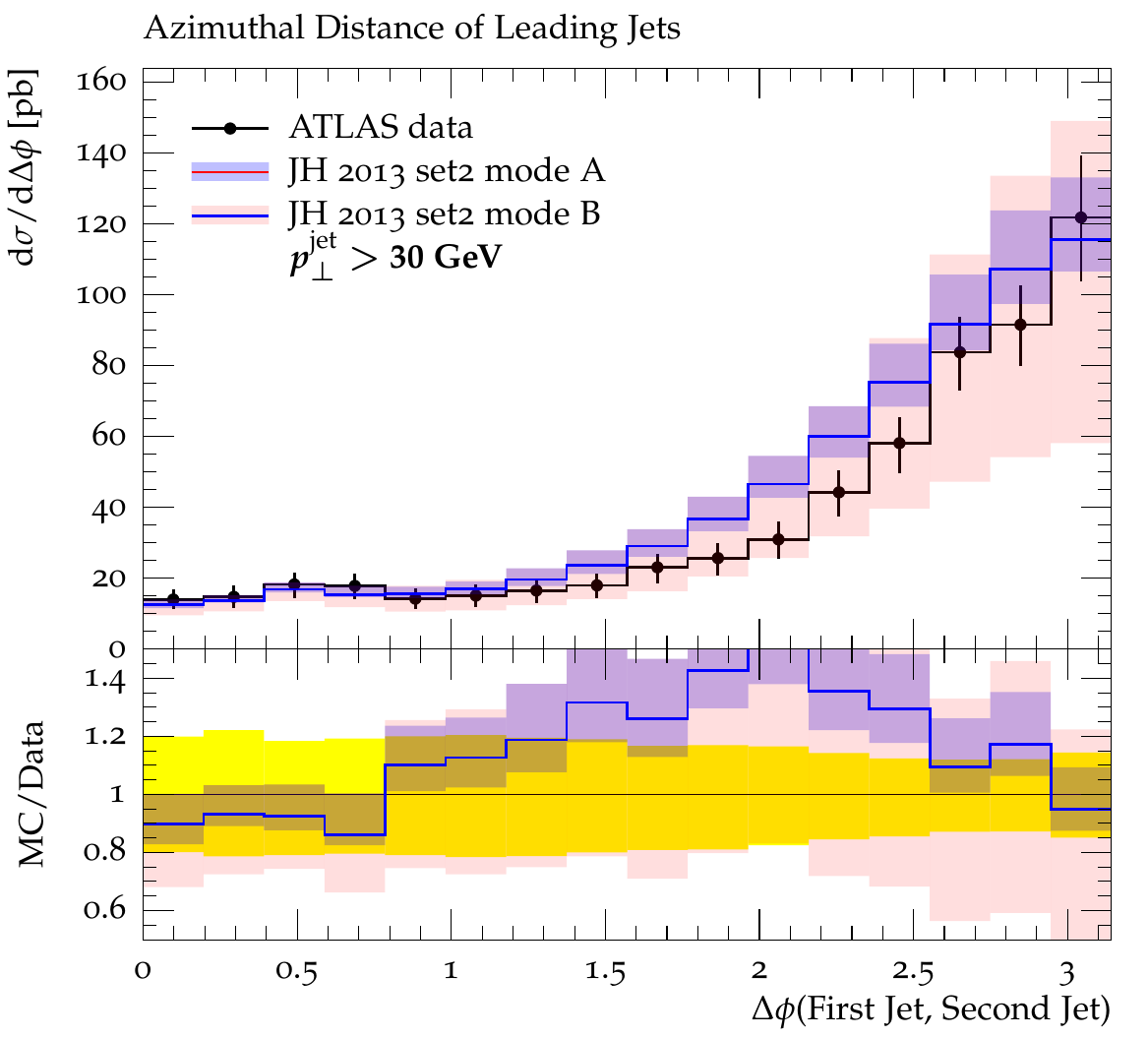}
\includegraphics[scale=0.67]{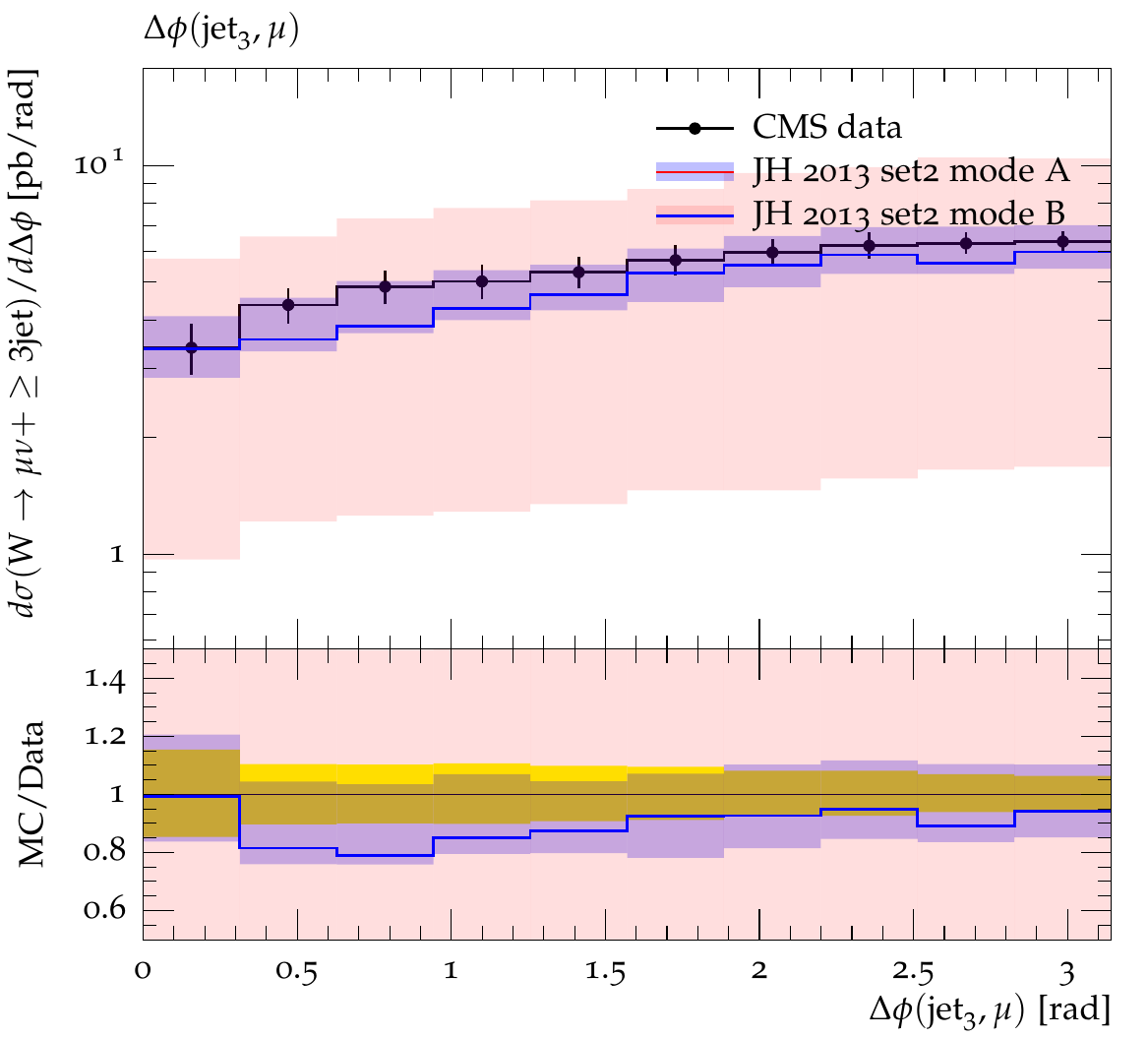}
  \caption{\it (left)  Azimuthal distance of the leading jets 
  associated with   
$W $-bosons; (right) azimuthal correlation of the 
third jet to the $W$.     The purple and pink bands 
correspond to mode A and mode B  
as described in the text. 
 The experimental data are 
 from~\protect\cite{atlas-w-jets} (left)   and~\protect\cite{cms-w-jets} (right), 
with the  
experimental uncertainty represented by  the yellow band.}
\label{fig:azim}
\end{figure}

In Fig.~\ref{fig:azim} we 
turn to angular correlations. We consider two examples: 
the distribution in the azimuthal 
separation $\Delta \phi$ between the two hardest 
jets (left);  the  correlation 
 of the third jet to the $W$-boson (right).  
As noted earlier,  predictions of the structure of  angular 
correlations are a distinctive feature of the 
TMD  exclusive formulation.    
 The 
 shape of the   
experimental measurements is  well described, within 
the   theoretical  uncertainties, both at large $\Delta \phi$  and   down to 
 the decorrelated, small-$\Delta \phi$ region.

In conclusion, this work  shows how exclusive evolution equations in 
QCD at high energies  can be used to take into account  QCD 
contributions to the production of electroweak bosons plus multi-jets 
due to finite-angle soft gluon radiation, and estimate the associated 
theoretical uncertainties.  This will be  relevant  both to  
precision  studies of Standard 
Model physics and to new physics searches 
for which vector boson plus jets are an important background. 

Unlike  traditional   approaches  to    electroweak  boson production       
 including    effects of   the initial state's 
transverse momentum  in the low-$p_\perp$ region, 
   the  formulation  of TMD pdfs and factorization   employed in 
  this work 
 incorporates physical effects which 
persist  at high $p_\perp$  
and  treats   final states of high   multiplicity. 
The effects studied   
come from  multiple gluon emission at finite angle and the 
associated color coherence~\cite{mw92,Catani:1989sg,hj-ang}, 
and are present  to all   orders 
 in the strong coupling $\alpha_s$.  In particular,  they 
are  beyond 
next-to-leading-order 
perturbation theory matched with collinear parton 
showers~\cite{hoeche13}. They can  contribute 
significantly to the estimate of theoretical uncertainties in 
multi-jet distributions at  high energies.

The method of this work incorporates the experimental information 
from the high-precision 
DIS combined measurements~\cite{comb-charm,Aaron:2009aa}. 
The use of the TMD density determined~\cite{hj-updfs-1312} from these 
measurements   in   the comparison 
with the  LHC  $W$ + $n$-jet   data  indicates  
that detailed features of the associated final states can be 
obtained  both  for the     leading jet 
 and  the  subleading jets. It    
  underlines the consistency  of the  
 physical picture  which can be extended 
from DIS   to Drell-Yan processes 
to describe  QCD multi-jet dynamics. 
It also  points to the relevance of Monte Carlo event 
generators which aim at   including  parton branching 
at  transverse momentum dependent level (see 
e.g.~\cite{jadach2012,Jung:2010si}).

Future applications may  employ  vector boson $ pp $ 
data to advance our knowledge of transverse momentum 
parton distributions~\cite{herafitter,mert-rog}.  
Vector boson plus jets are a benchmark process for 
QCD studies of multi-parton interactions~\cite{ellie}, 
  and may help shed light on topical issues 
 in the physics of  forward jet production~\cite{monika}. 
A  program  
 combining Drell-Yan and Higgs measurements 
can become  viable at high luminosity~\cite{cipriano}  
to carry out precision QCD  studies 
accessing gluon transverse momentum and polarization 
distributions~\cite{cipriano,dendunnen}.

\section*{Acknowledgments}
We thank D.~Baumgartel for help with 
the CMS preliminary {\sc Rivet}  plug-in.

\end{document}